\documentclass[twocolumn,aps,prl,superscriptaddress,notitlepage,longbibliography]{revtex4-2}

\usepackage{graphicx}
\usepackage{epsfig}
\usepackage{float}
\usepackage{dcolumn}
\usepackage{bm}
\usepackage{amsmath}
\usepackage{float}
\usepackage{amssymb}
\usepackage{lipsum}
\usepackage{titletoc}
\usepackage{makecell}
\usepackage[colorlinks,linkcolor=blue]{hyperref}
\usepackage{array}
\usepackage{multirow}
\usepackage{color}
\usepackage{bbding}
\usepackage{stackengine}

\newcommand{\be}{\begin{equation}}
\newcommand{\ee}{\end{equation}}

\begin{document}

\title{Topological metal and high-order Dirac point in cubic Rashba model}

\author{Haijiao Ji}
\affiliation{Harbin Institute of Technology, Shenzhen, 518055, P. R. China}
\author{Ning Zhang}
\affiliation{Harbin Institute of Technology, Shenzhen, 518055, P. R. China}
\author{Noah F. Q. Yuan}
\email{fyuanaa@connect.ust.hk}
\affiliation{Harbin Institute of Technology, Shenzhen, 518055, P. R. China}

\begin{abstract}
We investigate the properties of the two-dimensional model with Rashba-type spin-orbit coupling cubic in electron momentum.
In the normal phase, edge states emerge on open boundaries.
In the superconducting phase, edge states could evolve into gapped fermionic edge states.
Applications to realistic materials of interface superconductors are also discussed.
\end{abstract}
\maketitle

\textit{\textcolor{blue}{Introduction.}}---
The Rashba effect is one of the well-known manifestations of spin-orbit couplings (SOCs) in solids where inversion symmetry is broken by an electric field normal to the heterointerface \cite{Rashba0,Rashba}. Another well-known type of SOCs due to the inversion symmetry breaking is the so-called Dresselhaus SOC \cite{Dresselhaus}. In both cases, the SOC Hamiltonian is linear in electron momentum $\bm k$, and the electron spins exhibit one full winding in moving around the closed Fermi contour. Usually such $\bm k$-linear SOCs are leading order effects in inversion symmetry breaking systems \cite{klinear1,klinear2,klinear3}.

In certain systems the leading SOCs will not be $\bm k$-linear \cite{cubic0,cubic1,cubic2,cubic3,cubic4}. For example in transition metal dichalcogenides (TMDs) or the Kane-Mele model, due to the special point group symmetry, the leading order SOC near $\Gamma$ point is $\bm k$-cubic, while the electron spin is pinned to the out-of-plane direction. 

Other examples of nonlinear SOCs can be found in interfaces such as (111) surface of GaAs quantum well \cite{well0,well2}, (001) surface of oxide SrTiO$_3$ (STO) \cite{STO,STO1,STO2,STO3,STO4}, and Si-terminated (001) surface of the rare-earth antiferromagnet RRh$_2$Si$_2$ \cite{TRS,TRS1,TRS2} (R denotes the rare-earth element such as Yb, Tb, etc.), where the $\bm k$-cubic Rashba SOC can well describe the dominant band structure, as verified by experiments \cite{cubic1,TRS}. 
In GaAs quantum well, fine tuning is needed to cancel $\bm k$-linear Rashba and Dresselhaus effects \cite{cancel}.
When the dominant electrons are $d$- (STO) or $f$-orbitals (TbRh$_2$Si$_2$), and the leading order SOCs could become cubic \cite{STO,TRS2}.

Superconductivity has been found experimentally in some of the above systems with $\bm k$-cubic SOCs, including bulk and two-dimensional (2D) TMDs \cite{SC-TMD1,SC-TMD2,SC-TMD3,SC-TMD4,SC-TMD5}, LaAlO$_3$/SrTiO$_3$ (LAO/STO) interfaces \cite{SC-LAO,SC-STO,SC-STO1}, and RRh$_2$Si$_2$ \cite{SC-TRS}. Correspondingly, unconventional superconductivity such as Ising superconductivity \cite{ising,ising1,ising2} and topological nodal superconductivity \cite{TNSC,TNSC1,TNSC2} has been found in 2D TMDs. It is therefore interesting to investigate the properties of superconductors with $\bm k$-cubic SOCs such as LAO/STO interfaces and RRh$_2$Si$_2$. 

\begin{figure}
\includegraphics[width=\columnwidth]{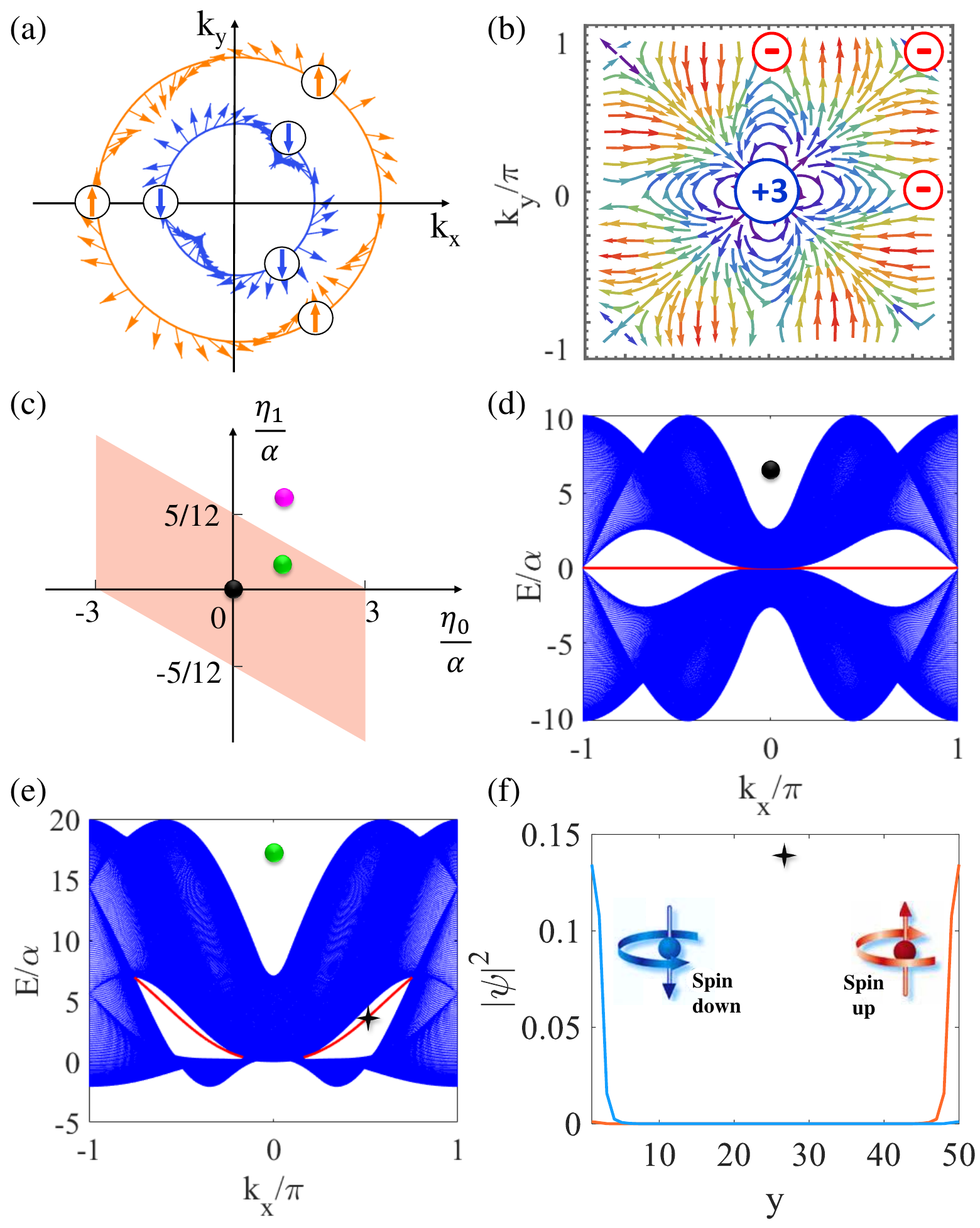}
\caption{(a) Spin texture of the Fermi contours of the cubic model Eq. (\ref{eq1}). electron spins have three full windings in moving around the Fermi contours.
(b) The distribution of spin field $(S_x,S_y)$ in the Brillouin zone, where the spin winding number $W=-3$ at $\Gamma=(0,0)$ point (colored blue), while $W=+1$ at $X=(\pi,0)$, $Y=(0,\pi)$ and $M=(\pi,\pi)$ points (colored red). The spin winding number $W$ is defined in Eq. (\ref{eq_W}) and the spin field $(S_x,S_y)$ is defined underneath.
(c) Phase diagram of the cubic Rashba model Eq. (\ref{eq1}), where edge states can be found explicitly in the pink region, for example the black (d) and green (e) dots, while edge states are mixed with bulk states in the white region such as the red dot [Fig. \ref{fig4}(a)]. (d) and (e) are corresponding energy spectra with open boundary conditions along $y$-direction with 200 lattice sites. (f) is the wavefunction of the edge states at momentum indicated by black dagger in (e) with 50 lattice sites.}\label{fig1}
\end{figure}

In this work, we focus on the abstract two-dimensional model equipped with the cubic Rashba SOC, and study the superconducting phase. In this cubic Rashba model, the SOC Hamiltonian is cubic in electron momentum $\bm k$, and the electron spins exhibit \textit{three} full windings in moving around the closed Fermi contour, in contrast to the one full winding of the linear Rashba case.
In the linear Rashba model, the so-called topological metal has been proposed to describe its normal phase \cite{Tometal,Tometal1,Tometal2,Tometal3}, and the corresponding superconducting phase will also be affected to host finite-momentum superconductivity.
In the cubic Rashba model, the notion of topological metal also applies to its normal phase, with edge states emerging on open edges. 
In the superconducting phase, these edge states could evolve into gapped fermionic edge states.
Such exotic behaviors may have potential impacts on the experimental measurements of superconductors such as LAO/STO interfaces and RRh$_2$Si$_2$.

\begin{figure}
\includegraphics[width=\columnwidth]{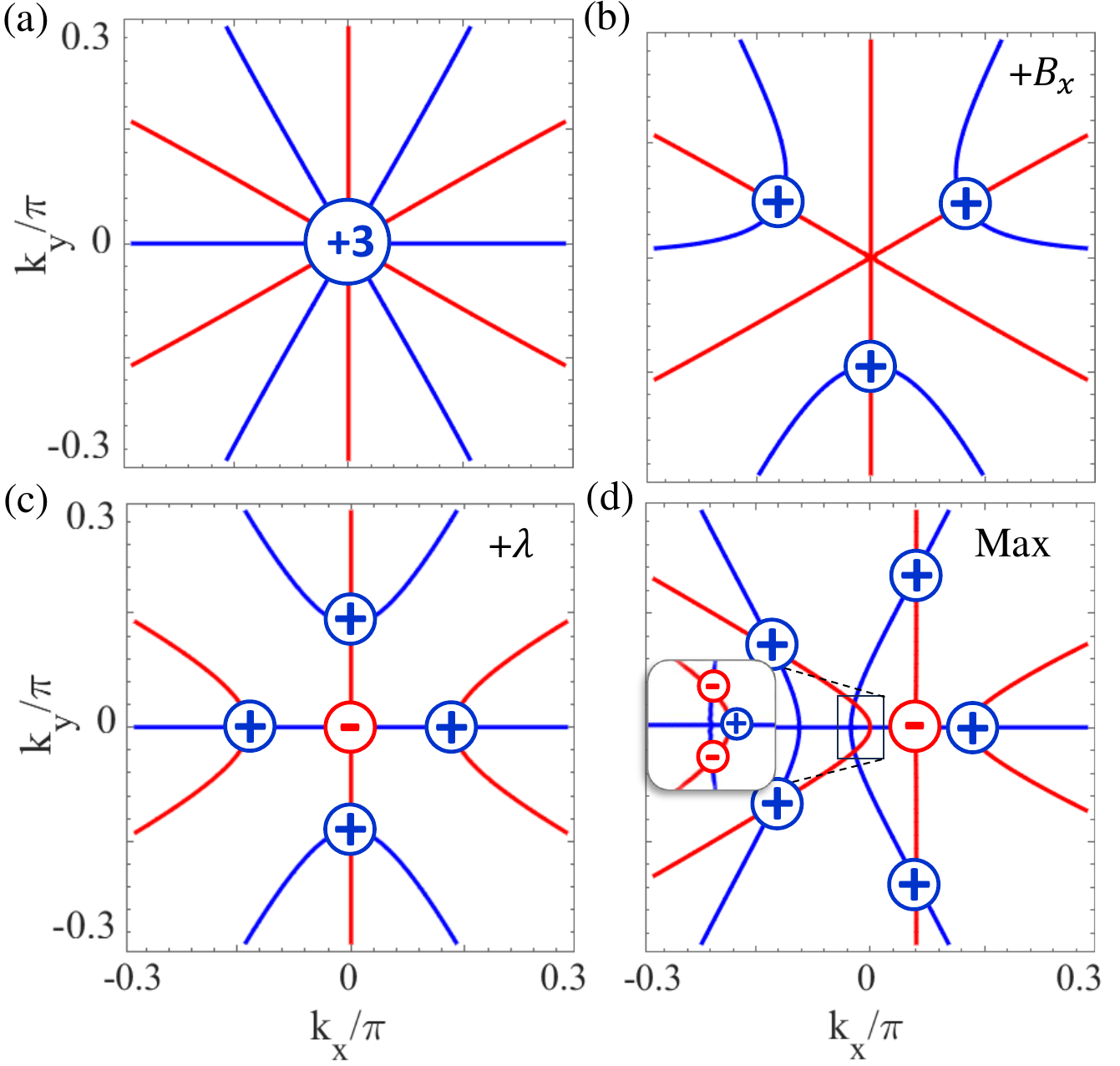}
\caption{Zero contours $S_x=0$ (blue) and $S_y=0$ (red) of the spin field $(S_x,S_y)$, and the intersecting points of blue and red lines are Dirac points with corresponding spin winding numbers. (a) One cubic Dirac point at zero field. (b) Three linear Dirac points under in-plane Zeeman field $B_x=0.1\alpha$ along $x$-axis. (c) Five linear Dirac points under linear Rashba $\lambda=0.1\alpha$. (d) Nine linear Dirac points under symmetry-breaking perturbations.}\label{fig2}
\end{figure}

\textit{\textcolor{blue}{Topological metal and edge states.}}---
It is theoretically proposed and experimentally verified that the following cubic Rashba model could describe the low-energy band structure of interface superconductor LAO/STO \cite{cubic1},
\begin{equation}
{H}_{\rm cR}(\mathbf{k}) =\eta_0 k^2+\eta_1 k^4 +\frac{1}{2}i\alpha(k_{+}^3\sigma_{-}-k_{-}^3\sigma_{+})
 \label{eq1}
\end{equation}
with electron momentum $\mathbf{k}=(k_x,k_y)$, $k=|\mathbf{k}|$, $k_{\pm}=k_x\pm ik_y$, $\sigma_{\pm}=\sigma_x\pm i\sigma_y$, and Pauli matrices $\bm\sigma$ denoting spin. 
Here $\eta_{0,1}$ are coefficients for kinetic terms, and $\alpha$ is the cubic Rashba SOC coefficient.

The unitary symmetries of the cubic Rashba model form the point group $D_{2d}$ with three generators:
Twofold in-plane rotation $C_{2z}$, 
twofold rotation $C_{2x'}$ along the diagonal direction $\bm x'\equiv(\hat{\bm x}+\hat{\bm y})/\sqrt{2}$, and in-plane mirror reflection $M_x: (k_x,k_y)\to(-k_x,k_y)$, $(\sigma_x,\sigma_y,\sigma_z)\to(\sigma_x,-\sigma_y,-\sigma_z)$.
Though the point group is anisotropic, the energy bands of the cubic Rashba model are isotropic
\begin{equation}
E_{\pm}(\mathbf{k})=\eta_0 k^2+\eta_1 k^4\pm\alpha k^3,
\label{eq2}
\end{equation}
and the two Fermi contours are hence circular and concentric. The inner $(+)$ and outer $(-)$ Fermi contours correspond to the upper $(+)$ and lower $(-)$ bands.

Due to the cubic Rashba SOC, electron spins have three full windings in moving around the Fermi contours as shown in Fig. \ref{fig1}(a), which can be characterized by the Berry phase $\pi$ of both Fermi contours as enforced by the parity symmetries including time-reversal symmetry $\mathcal{T}$, mirror symmetries $M_x$, $M_y$ and combined symmetry $C_{2z}\mathcal{T}$. The metallic phase with quantized nonzero Berry phase $\pi$ of the Fermi contours is known as the topological metal \cite{Tometal,Tometal1,Tometal2,Tometal3}, which is consistent with quantum oscillation data of LAO/STO interfaces \cite{STO3}.

In the topological metal, there are odd numbers of Dirac points enclosed by the Fermi contours. However, the detailed properties of the Dirac point are not captured by the Berry phase, hence the properties of the topological metal is yet to be revealed.

For this purpose, we introduce the spin winding number to describe the topological property of the Dirac point
\begin{eqnarray}\label{eq_W}
    W=\frac{1}{2\pi}{\rm Im}\oint_{\rm FS}\frac{dS_x+idS_y}{S_x+iS_y},
\end{eqnarray}
where FS denotes the Fermi surface (or Fermi contour in 2D) and $\bm S=(S_x,S_y,S_z)=\langle\bm\sigma\rangle$ is the spin expectation value of the lower band $E_{-}$.
The spin winding number of the cubic Dirac point in Eq. (\ref{eq1}) can be worked out as $W=3$. 
For linear Dirac point $H=\sum_{ij} k_iA_{ij}\sigma_j$, the spin winding number is $W={\rm sgn}(\det A)$.

The cubic Rashba model Eq. (\ref{eq1}) is defined near the $\Gamma$ point.
When this model is extended to the entire Brillouin zone (BZ), the Poincar\'e-Hopf index theorem dictates that the total spin winding number is zero
\begin{equation}
    \sum_i W_i=0.
\end{equation}
The BZ compatible with point group $D_{2d}$ would be that of a square lattice. Without loss of generality, we assume the lattice constant of the square lattice to be 1.
When the time-reversal-invariant points $X=(\pi,0)$, $Y=(0,\pi)$ and $M=(\pi,\pi)$ are linear Dirac points $|W_{X,Y,M}|=1$ while $\Gamma$ is cubic $W_{\Gamma}=3$, the only choice of spin winding numbers is $W_{X,Y,M}=-1$ due to the Poincar\'e-Hopf index theorem, which is shown in the distribution of spin field $(S_x,S_y)$ in the BZ in Fig. \ref{fig1}(b). Details of the tight-binding cubic Rashba model of Eq. (\ref{eq1}) can be found in the Appendix.

The nontrivial spin winding number of the cubic Dirac point indicates the nontrivial bulk topology and hence topological edge states in the tight-binding model. In the phase diagram of Fig. \ref{fig1}(c), when the parameters $\eta_{0,1}/\alpha$ fall in the topological region (colored pink), the two energy bands are separated by local gaps as shown in Fig. \ref{fig1}(d) and (e). Under the open boundary condition of $x$ or $y$-direction, edge states emerge, whose dispersions are within local gaps [red lines in Fig. \ref{fig1}(d) and (e)], and whose wavefunctions are localized on two edges with opposite spin polarization along $z$-axis as shown in Fig. \ref{fig1}(f). Notice that the edge states are spin degenerate. 

To figure out the origin of such edge states, we can consider the chiral limit $\eta_0=\eta_1=0$ as shown in Fig. \ref{fig1}(d). In this chiral limit, the cubic model Eq. (\ref{eq1}) has the chiral symmetry $\{H_{\rm cR},\sigma_z\}=0$ and hence hosts zero energy modes on its open boundaries. Classified in the BDI class, nodal points of the bulk energy spectrum are characterized by the spin winding number $W$, and under open boundary condition, flat bands will emerge between projected nodal points with opposite spin winding numbers. 
In particular, when $y$-direction is open, the nodal points are projected to $k_x=0$ with net spin winding number $W(k_x=0)=+3-1=+2$, and $k_x=\pm\pi$ with net spin winding number $W(k_x=\pm\pi)=-1-1=-2$. Thus a doubly degenerate flat band emerges connecting $k_x=0$ and $k_x=\pi$, and another doubly degenerate flat band emerges connecting $k_x=0$ and $k_x=-\pi\equiv\pi({\rm mod}\ 2\pi)$.
Furthermore, zero-energy edge states are eigenstates of chiral symmetry $\sigma_z$, and are hence spin polarized along $z$-axis. As shown in Fig. \ref{fig1}(f), we hence find spin-degenerate edge states.

Away from the chiral limit, one may treat $\eta_{0,1}$ as perturbations, under which the nodal points evolve into Dirac points connecting upper and lower bands, and the edge states become dispersive. As long as the local gaps are not closed, the corresponding edge states are protected by the bulk topology and separated from bulk states. As a result, we obtain phase diagram of Fig. \ref{fig1}(c).

\textit{\textcolor{blue}{Topological metal under perturbations.}}---
We consider the following perturbations in the metallic phase
\begin{equation}
{H}_{\rm per}(\mathbf{k}) =\frac{1}{2}i\lambda(k_{+}\sigma_{+}-k_{-}\sigma_{-})+\bm B\cdot\bm\sigma
 \label{eq3}
\end{equation}
where $\lambda$ is the linear Rashba coefficient and $\bm B$ is the Zeeman field. These perturbations are allowed under point group $D_{2d}$, under which the number and positions of Dirac points of the band structure will change.

Under an in-plane Zeeman field, the single cubic Dirac point in Fig. \ref{fig2}(a) is split into three linear Dirac points as shown in Fig. \ref{fig2}(b), forming a threefold structure. Under the linear Rashba effect, the cubic Dirac point is split into five linear Dirac points as shown in Fig. \ref{fig2}(c), forming a fourfold structure. During the Dirac point splitting process, the total spin winding number in the neighborhood of the $\Gamma$ point is conserved due to topology \cite{class}, as shown in Fig. \ref{fig2}(b) and (c).
In fact, the two energy bands can be expressed in terms of the polar coordinates of the electron momentum $\mathbf{k}=k(\cos\theta,\sin\theta)$ as 
\begin{eqnarray}
    E_{\pm}(\mathbf{k})=\eta_0 k^2+\eta_1 k^4\pm D(k,\theta),
\end{eqnarray}
where the energy bands are threefold symmetric under an in-plane Zeeman field $\bm B=B(\cos\phi,\sin\phi)$, while fourfold symmetric under linear Rashba effect $\lambda$:
\begin{equation}
    D(k,\theta)=
    \begin{cases}
        \sqrt{B^2+\alpha^2k^6-2B\alpha k^3\sin(3\theta-\phi)} & \lambda=0,\\
        k\sqrt{\lambda^2+\alpha^2k^4-2\lambda\alpha k^2\cos 4\theta } & B=0.
    \end{cases}
\end{equation}
With both in-plane Zeeman field and linear Rashba effect, the in-plane rotation symmetry of the energy bands is then lost.

We introduce the multiplicity (also known as the Milnor number) $\mu$ of the Dirac point, which is defined as the maximal number of split Dirac points under perturbations \cite{class}.
Here perturbations are defined as the Hamiltonian terms with lower orders in momentum than that of the Dirac point.
The cubic Dirac point can be split into at most five linear Dirac points under the symmetry-allowed perturbations in Eq. (\ref{eq3}), hence the symmetry-constraint multiplicity of the cubic Dirac point is $\mu_{\rm sym}=5$.
Without symmetry constraints, the cubic Dirac point can be split into at most $\mu=W^2=9$ linear Dirac points, as shown in Fig. \ref{fig2}(d) where the time-reversal symmetry $\mathcal{T}$ is broken and the point group is reduced to $C_{1v}=\{1,M_y\}$.

\begin{figure}
\includegraphics[width=\columnwidth]{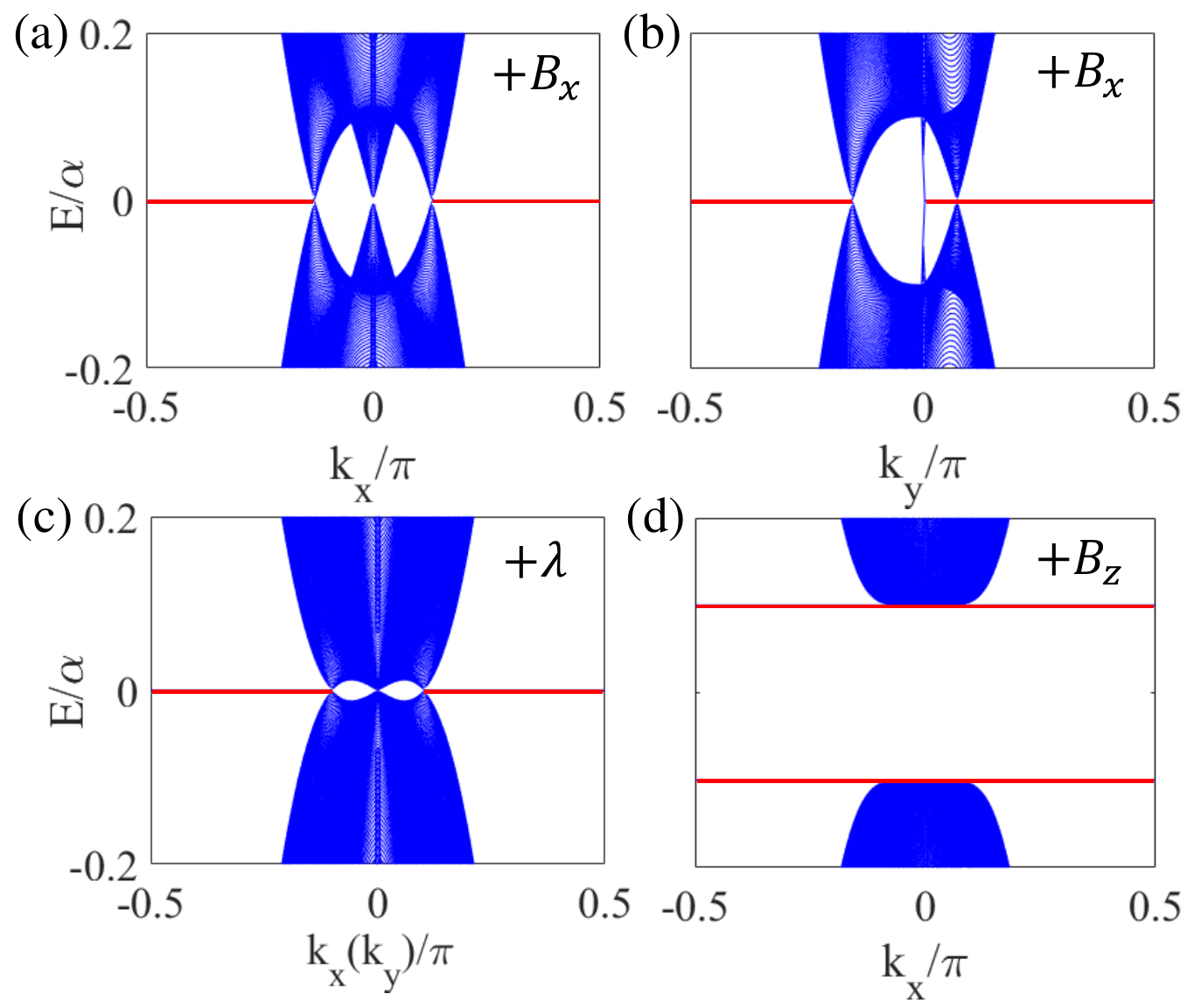}
\caption{Energy spectra near $\Gamma$ point in the chiral limit under external Zeeman fields (a,b,d) and linear Rashba effect (c), where edge states are in red color. Parameters are the same as Fig. \ref{fig2}, and $B_z=0.1\alpha$ in (d).}\label{fig3}
\end{figure}

We have discussed the effects of perturbations on the bulk spectrum especially the Dirac points. Due to the bulk-edge correspondence, the edge states will also be affected. The general mechanism indicates that edge states will emerge between projected nodal points with opposite spin winding numbers. 
For simplicity we consider the chiral limit first.
As shown in Fig. \ref{fig3}, edge states are found under open boundary conditions, which are flat bands in the chiral limit. Notice that under in-plane Zeeman field (a, b) and linear Rashba effect (c), the net spin winding numbers of the nodal points near projected $\Gamma$ point can become 0, $\pm 1$ besides $\pm 2$. Notice that under an out-of-plane field $B_z$ (d), the spin-degenerate edge band will get split into two spin polarized bands, whose wavefunctions are similar to those shown in Fig. \ref{fig1}(f).

The introduction of kinetic terms $\eta_{0,1}$ will gradually close the local gaps, which disperse and eventually diminish edge states. In the following, we would like to show that even when kinetic terms close the local gaps and diminish edge states in the normal phase, fermionic edge states will get revealed under appropriate pairing potentials in the superconducting phase.

\textit{\textcolor{blue}{Topological metal under pairings.}}---
We consider the cubic Rashba model with pairing potential, 
whose Bogouliubov-de Gennes (BdG) Hamiltonian reads
\begin{equation}\label{eq_BdG}
{H}_{\rm B d G}(\mathbf{k}) =
\left(\begin{array}{cc}
{H}_{\rm cR}(\mathbf{k})-\varepsilon_F & H_{\Delta}(\mathbf{k}) \\
 H_{\Delta}^{\dagger}(\mathbf{k})  & -{H}^{*}_{\rm cR}(-\mathbf{k})+\varepsilon_F
\end{array}\right),
\end{equation}
where $\varepsilon_F$ is the Fermi energy and the generic pairing is
\begin{eqnarray}\label{eq_D0}
    H_{\Delta}=(\psi+\bm d\cdot\bm\sigma)i\sigma_y,
\end{eqnarray}
with the spin-singlet $\psi(\mathbf{k})$ and spin-triplet $\bm d(\mathbf{k})$ pairings.
Due to the Pauli exclusion principle, the spin-singlet is even in momentum $\psi(-\mathbf{k})=\psi(\mathbf{k})$ and spin-triplet is odd $\bm d(-\mathbf{k})=-\bm d(\mathbf{k})$. According to the symmetry group $D_{2d}$, the possible spin-singlet and spin-triplet pairings are calculated as shown in Table. \ref{table1}.

\begin{figure}
\includegraphics[width=\columnwidth]{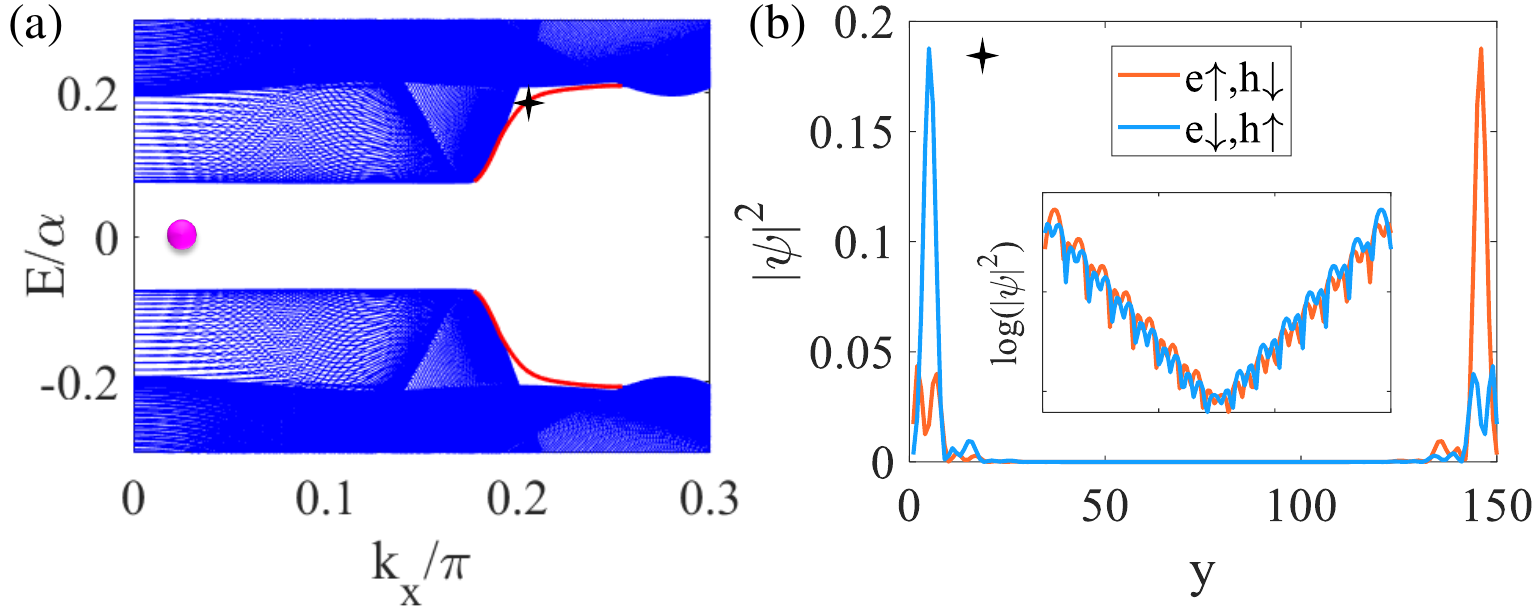}
\caption{(a) Superconducting spectrum of the BdG Hamiltonian Eq. (\ref{eq_BdG}) with $\psi=\Delta_0+\Delta_1 k^2$ and $\bm d=\bm 0$. The red lines denote the gapped fermionic edge states. Parameters are $(\eta_0,\eta_1)=(1,\frac{1}{2})\alpha$ [red dot in Fig. \ref{fig1}(c)] and $(\Delta_0,\Delta_1)=(0,\frac{1}{4})\alpha$. (b) Edge state wavefunctions at the black dagger in (a), with electron/hole (e/h) and spin ($\uparrow/\downarrow$) components. Inset: Logarithm of edge state wavefunctions, showing the exponential localization of edge states.}\label{fig4}
\end{figure}

Under pairings, edge states will be revealed, as long as
\begin{eqnarray}\label{eq_D}
    \Delta_{+}<\Delta_{-}
\end{eqnarray}
where $\Delta_{\pm}$ denote the pairing gaps of the inner $(+)$ and outer $(-)$ Fermi contours respectively.
For example, one can consider the $s$-wave pairing with constant $(\Delta_0)$ and extended $(\Delta_1)$ parts
$\psi = \Delta_0+\Delta_1 k^2$.
As shown in Fig. \ref{fig4}(a), when $\Delta_1>0$ and the condition Eq. (\ref{eq_D}) is satisfied, the edge states (colored red) will emerge joining the inner $(+)$ and outer $(-)$ Fermi contours with pairing gaps $\Delta_{+}$ and $\Delta_{-}$ respectively, whose wavefunction is also localized on two open edges with opposite spin polarization along $z$-axis as shown in Fig. \ref{fig4}(b), where electron spin $\bm\sigma$ is equivalent to hole spin $-\bm\sigma^*$.
Importantly, the condition Eq. (\ref{eq_D}) applies when bulk and edge states are separated [Fig. \ref{fig1}(c) pink region] or mixed [Fig. \ref{fig1}(c) white region]. 

\begin{table}[ht]
\centering
\caption{\textbf{Possible pairings in the cubic Rashba model.}}
\begin{center}  
\begin{tabular}{c|c|c}  
\hline  
$D_{2d}$ & $\psi$ & $\bm d$\\
\hline 
\multirow{3}*{$A_{1}$} & $1$ & $k_y\hat{\bm x}+k_x\hat{\bm y}$ \\
& $k^2$ & $k_x^2k_y\hat{\bm x}+k_y^2k_x\hat{\bm y}$ \\
& $k^4,{\rm Re}k_{+}^4$ & $k_y^3\hat{\bm x}+k_x^3\hat{\bm y}$ \\
\hline
\multirow{3}*{$A_{2}$} & \multirow{3}*{${\rm Im}k_{+}^4$} & $k_x\hat{\bm x}-k_y\hat{\bm y}$ \\
&  & $k_xk_y^2\hat{\bm x}-k_yk_x^2\hat{\bm y}$  \\
&  & $k_x^3\hat{\bm x}-k_y^3\hat{\bm y}$ \\
\hline
\multirow{3}*{$B_{1}$} & \multirow{3}*{${\rm Im}k_{+}^2$} & $\bm k$ \\ 
&   & $k_xk_y^2\hat{\bm x}+k_yk_x^2\hat{\bm y}$  \\
&   & $k_x^3\hat{\bm x}+k_y^3\hat{\bm y}$ \\
\hline
\multirow{3}*{$B_{2}$} & \multirow{3}*{${\rm Re}k_{+}^2$} & $\hat{\bm z}\times\bm k$ \\ 
&   & $k_x^2k_y\hat{\bm x}-k_y^2k_x\hat{\bm y}$ \\
&   & $k_y^3\hat{\bm x}-k_x^3\hat{\bm y}$ \\
\hline
{$E$} & {0} &{$\left(\begin{array}{c}
k_x\hat{\bm z} \\
k_y\hat{\bm z}
\end{array}\right)$, $\left(\begin{array}{c}
k_x k_y^2\hat{\bm z} \\
k_x^2 k_y\hat{\bm z}
\end{array}\right)$, $\left(\begin{array}{c}
k_x^3\hat{\bm z} \\
k_y^3\hat{\bm z}
\end{array}\right)$}\\
\hline
\end{tabular}  
\end{center} 
\label{table1}
\end{table}

\textit{\textcolor{blue}{Discussions.}}---
As experimental candidates, we mainly consider LAO/STO interface \cite{cubic1}. 
The conduction band of bulk SrTiO$_3$ originates from Ti 3$d$ orbitals, whose bottom is composed of $t_{2g}$ states ($d_{xy},d_{yz}$ and $d_{xz}$-like orbitals) due to crystal field splitting \cite{DFT}. Including the atomic SOC, the lowest energy states for bulk SrTiO$_3$ consist of fourfold degenerate bands, which correspond to total angular momentum $J=3/2$ labeled by $z$-component $J_z=\pm 1/2$ and $\pm 3/2$. Due to the interface potential induced by LaAlO$_3$, it turns out that $d_{xy}$-like orbitals ($J_z=\pm 3/2$) dominate the bottom conduction bands of LAO/STO interface. 
Under point group $D_{2d}$ on the basis of $J_z=\pm 3/2$ states, the cubic Rashba model Eq. (\ref{eq1}) could be derived, and the cubic Rashba coefficient $\alpha$ in Eq. (\ref{eq1}) can be tuned by the electric gate voltage \cite{cubic1}.

The edge states and bulk states can mix at the same energy, which prevents differentiating bulk and edge states in the topological metallic phase.
One possible way is to apply an out-of-plane magnetic field to create Landau levels, so that bulk states become gapped. 
Another way is to consider the superconducting phase under magnetic field. In that case, the edge states and bulk states give rise to different spatial profiles of supercurrent density. To be more precise, one may prepare a Josephson junction in terms of LAO/STO, and measure the Fraunhofer pattern under an out-of-plane field. By appropriate transforms of the Fraunhofer pattern \cite{DF}, the supercurrent density distribution can be obtained, where the edge localized supercurrent corresponds to edge states \cite{Wang}.



\section{Acknowledgements}
We thank Dong-Hui Xu for bringing us the cubic Rashba model.
We thank K. T. Law, B. Q. Lv and Yuanfeng Xu for helpful discussions.
This work is supported by the National Natural Science Foundation of China (Grant. No. 12174021).

\section*{Appendix: Tight-binding cubic Rashba model}
The continuum cubic Rashba model reads
\begin{equation}
{H}_{\rm cR}(\mathbf{k}) =\eta_0 k^2+\eta_1 k^4 +\frac{1}{2}i\alpha(k_{+}^3\sigma_{-}-k_{-}^3\sigma_{+}),
\end{equation}
where
\begin{equation}
k^2=k_x^2+k_y^2,\quad
k^4=(k_x^2+k_y^2)^2,
\end{equation}
\begin{equation}
{k}_{+}^3=(k_x+ik_y)^3=k_x^3+3ik_x^2k_y-3k_xk_y^2-ik_y^3,
\end{equation}
\begin{equation}
{k}_{-}^3=(k_x-ik_y)^3=k_x^3-3ik_x^2k_y-3k_xk_y^2+ik_y^3.
\end{equation}
Via the following substitution
\begin{equation}
\begin{split}
&k_{x(y)}\to\sin k_{x(y)},\\
&k_{x(y)}^2\to 2(1-\cos k_{x(y)}),\\
&k_{x(y)}^3\to 2\sin k_{x(y)}(1-\cos k_{x(y)}).\\
\end{split}
\end{equation}
The tight-binding Hamiltonian becomes
\begin{equation}
\begin{split}
H&=-2 \eta_0(\cos {k_x}+\cos k_y-2)+4 \eta_1(\cos k_x +\cos  k_y -2)^{2}\\
&+\alpha(3 \cos  2k_y-4\sin k_x-\sin 2 k_x)\sigma_y\\
&-2\alpha \sin k_y(-3 \cos k_x+\cos k_y+2)\sigma_x.
\end{split}
\end{equation}

\end{document}